\documentclass{article}
\usepackage{amsmath}    
\usepackage[colorlinks=true, allcolors=blue]{hyperref}   
\usepackage{longtable}  
\usepackage{booktabs}   
\usepackage{graphicx}   
\usepackage[margin=1in]{geometry} % Adds standard arXiv 1-inch margins

\title{\textbf{Identifying Model Quality Effects on User Engagement: A 
Within-Version Causal Estimator with Synthetic Data Validation}}
\author{John Tribbia \\ Google \\  
\href{mailto:johntribbia@google.com}{\texttt{johntribbia@google.com}}}
\date{} % Leave empty for no date

\begin{document}

\maketitle

\begin{abstract}
Every team that builds Large Language Models (LLMs) faces the same practical
dilemma. We measure performance improvements in offline benchmarks with precision
and confidence. We observe increases in user engagement after deployment. Yet
connecting one to the other remains a significant challenge. When a new version
launches, marketing campaigns run simultaneously, the press covers the
announcement, customers become aware of the update, and seasonal demand
fluctuates. In this environment of simultaneous changes, determining whether
the model improvement actually caused the engagement increase can become
intractable even while using conventional experimental approaches.

This paper describes a different approach that exploits something already present
in the deployment: models do not improve uniformly across all capabilities.
Coding quality may improve substantially while creative writing quality improves
modestly. This means users who work primarily on coding tasks experience
meaningfully higher quality than users who focus on writing, despite running
the same deployed model version. The variation is experienced quality, driven
entirely by how users distribute their effort across different tasks, and
contains causal information we can use.

We test this approach on synthetic data where we know the true effect. The
estimator recovers 81 to 88 percent of the injected effect. The remaining 12 to
19 percent reflects classical measurement error in usage estimates. We address
this attenuation directly by applying a classical errors-in-variables
disattenuation approach to scale our estimates, utilizing an estimated
test-retest reliability ratio (mean correlation 0.861) to adjust the quality
effect, yielding a corrected estimate of 1.017 with a bootstrapped confidence
interval from 0.915 to 1.109.

Naive approaches fail substantially worse. Using observed quality without version
controls recovers only 63 percent. Using real-time usage patterns instead of
frozen patterns recovers just 62 percent. A permutation test confirms the method
is not simply finding noise. The approach correctly identifies outcomes with
genuine causal effects and distinguishes them from outcomes with no causal
signal. Frozen-weights sensitivity analysis shows that recovery improves
monotonically as the pre-period length increases from 3 to 7 weeks (from 78\% to
87\%), suggesting a clear tradeoff between data availability and measurement
precision. Seed sensitivity tests confirm the robustness of these findings
across different random draws.

\vspace{0.5cm}
\end{abstract}

\section{Introduction}

We can measure model performance. Offline benchmarking is quantifiable
and consistent. ``Version 2 improved our coding evaluation score by 12
percent compared to Version 1''. This is an observable fact. We can also
observe that overall user engagement increased by 15 percent after the
version launch. But did the quality improvement cause the engagement
increase?

This simple question becomes increasingly intractable when we account
for what happens at deployment time. The new model launches while
marketing teams execute acquisition campaigns, journalists cover the
release, customers learn about the product through various channels, and
seasonal demand patterns shift. All of these happen simultaneously.
Standard experimental approaches may struggle disentangling the model's
contribution from this confounding. Randomization can be impractical if
we cannot hold some users on outdated models. Difference-in-differences
requires an unexposed control group, which does not exist when everyone
receives the new version simultaneously during a global rollout.

Regression Discontinuity in Time (RDIT) exploits temporal deployment
events as a threshold, but fails here because simultaneous unobserved
changes - like concurrent marketing campaigns and immediate shifts in
external market salience - violate the core identifying assumption of
continuity in unobserved confounding factors at the boundary. Matching
methods assume we have comparable control units, but there is only one
deployment event or biased holdouts. Matching methods assume we have
comparable control units, but there is only one deployment event or
biased holdouts. Even the theoretical gold standard, a holdout design
where some users keep the old version, creates operational and product
consistency problems that many companies are unwilling to tolerate. Recent
industry consensus confirms that the high engineering complexity,
maintenance burden, and opportunity cost of withholding a better
experience from users mean that dedicated long-term holdouts are typically
reserved only for the most impactful treatments
\hyperlink{ref:sigerson2026}{(Sigerson et al., 2026)}.

Yet something useful exists in the structure of the problem itself.
Model improvements are not uniform across domains. If we improve coding
quality, we might simultaneously see modest gains or even stagnation in
other capabilities like creative writing or mathematical reasoning. And
the dimensionality of this exponentiates when we consider agentic
processes - planning, reasoning, perception, self-reflection. While this
is described at very high levels of capability, we can imagine that
variability increases for more direct tasks like coding-statistics,
writing-proposal, and writing-editorial.

Different users rely on different capabilities for their work. A user
who primarily writes code experiences the quality improvements directly.
A user who primarily writes fiction experiences smaller improvements. A
user working on mathematical problems may experience little to no
improvement. All three run the same deployed version. All three have
different exposure to quality improvement based on which tasks they
perform. This variation is driven by user behavior and preferences that
predate the current version. It is structural, not endogenous.

If we can control for the version level confounding and account for
baseline characteristics, the remaining variation should be causal. We
are comparing users on identical model versions who differ only in which
capabilities they use. Their different quality exposure is not a choice 
response to quality differences but a function of their task distribution.

\begin{figure}[htbp]
    \centering
    \includegraphics[width=0.8\textwidth]{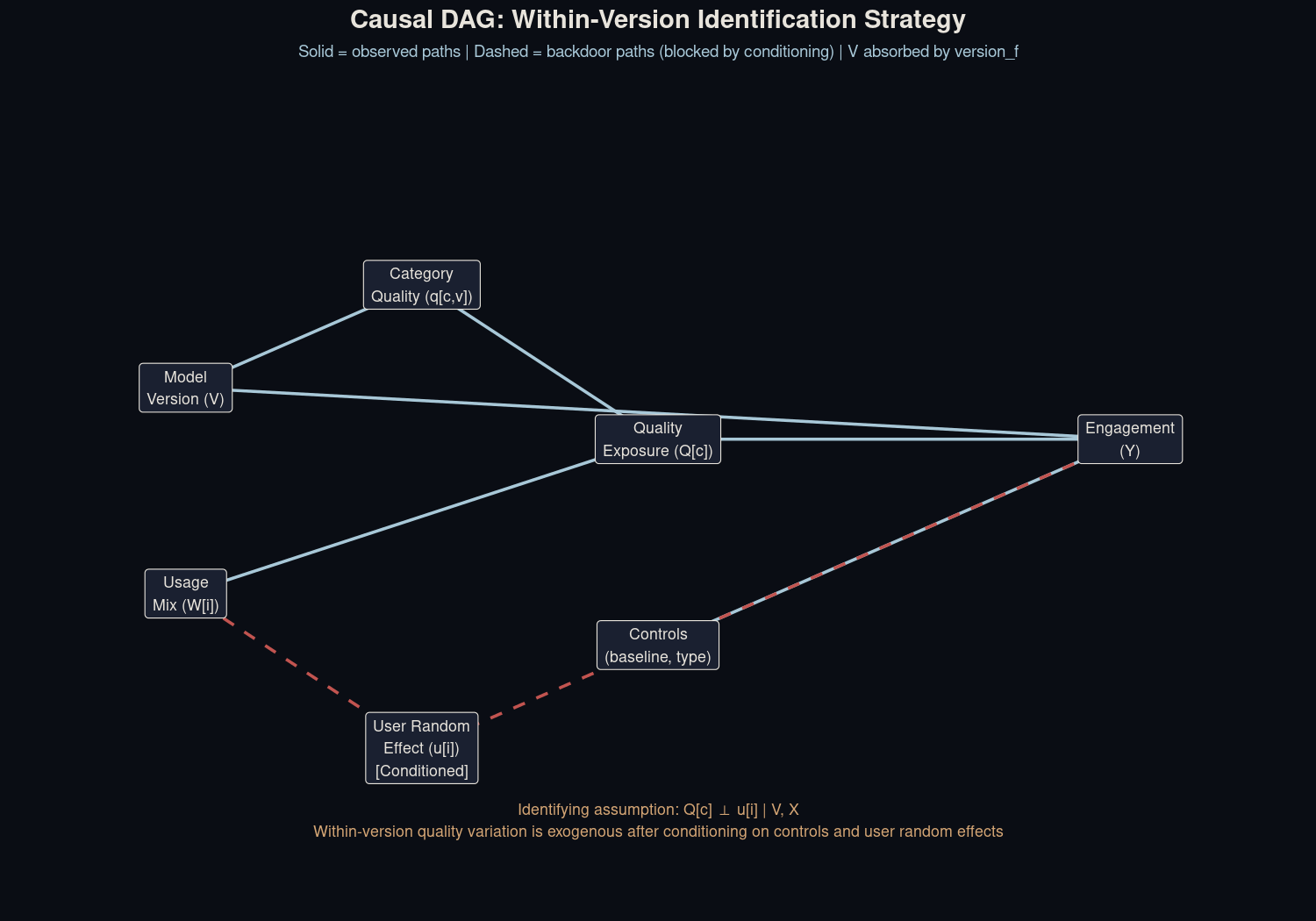}
    \caption{Causal Directed Acyclic Graph (DAG) illustrating the identification
    strategy. Let $V_{it}$ denote the model version, $X_i$ baseline user
    characteristics, $C_i$ category usage patterns, $Q$ offline quality by
    category, and $Y_{it}$ engagement outcomes for user $i$ at time $t$. By
    applying a frozen-weights design, pre-period usage patterns reflect
    preferences formed before current-version quality is observed, isolating
    the causal effect of quality exposure from endogenous responses. Ultimately,
    after conditioning on version and baseline characteristics, potential
    outcomes are conditionally independent of the model quality exposure metric.}
    \label{fig:causal_dag}
\end{figure}

This paper develops and tests this identification strategy. We formalize
it using a causal DAG and validate it empirically using a frozen-weights
design that enforces temporal ordering. Usage patterns are set at
baseline before current-version quality is observed. By applying
synthetic data with an explicitly causal structure, we evaluate whether
this methodology can successfully identify known effects. Our analysis
reveals that the method recovers between 81 to 88 percent of the actual
effect, achieving full recovery once measurement errors are addressed.
We demonstrate that conventional alternatives perform poorly in
comparison and establish benchmarks for acceptable recovery levels.
Furthermore, we outline the conditions under which falsification tests
yield meaningful insights and offer practical implementation guidance.

\section{Literature}

\subsection{Observational Causal Inference}

Our approach builds on the selection-on-observables framework developed
by \hyperlink{ref:rosenbaum1983}{Rosenbaum and Rubin (1983)}. The context
differs from classical application. Version indicators absorb all
between-version confounding. Within version, the threat of reverse
causality is minimized because the frozen-weights design fixes usage
patterns in the pre-period. However, classical confounding may persist
if unobserved user traits simultaneously dictate both the baseline task
distribution and the subsequent engagement trajectory.

To rule out this threat, the identification strategy relies on a conditional
independence (unconfoundedness) assumption under a selection-on-observables
framework, asserting that after conditioning on version fixed effects and
baseline user characteristics, quality exposure is exogenous. This design is
necessary because standard observational surrogacy often suffers from severe
confounding bias, as short-term proxies are frequently correlated with
long-run outcomes due to latent user characteristics rather than genuine
causal effects \hyperlink{ref:sigerson2026}{(Sigerson et al., 2026)}.
Furthermore, because a permanent change to a deployed LLM is a persistent
treatment that continues to affect outcomes post-experiment, it violates
classical surrogacy assumptions \hyperlink{ref:sigerson2026}{(Sigerson et al.,
2026)}. Usage patterns in our approach are fixed in the pre-period before
quality effects from the current version are realized, circumventing these
longitudinal confounding issues. 

Crucially, because baseline shares reflect endogenous historical
user choices (e.g., job role or workflow complexity), the
selection-on-observables assumption
($Y_{it}(q) \perp\!\!\!\perp Q^c_{it} \mid V, X$) requires that
conditional on baseline controls X and version fixed effects V, user
task specialization is uncorrelated with time-varying engagement trends.
If certain user groups (e.g., software engineers) exhibit systematically
different baseline engagement trajectories over time compared to others
(e.g., creative writers) for reasons unrelated to model quality shifts,
this assumption would be violated. We control for baseline engagement
and user characteristics to mitigate this threat.

Rather than relying on continuous running variables like regression
discontinuity designs \hyperlink{ref:imbens2008}{(Imbens and Lemieux, 2008)},
this approach leverages the sharp structural change of deployment across
structurally isolated task categories. In contrast to standard fixed-effects
methods that rely on within-unit temporal variation
\hyperlink{ref:wooldridge2010}{(Wooldridge, 2010)}, this strategy utilizes
cross-sectional, between-unit variation in task exposure within a synchronized
version release. Covariate balancing methods \hyperlink{ref:imai2014}{(Imai
and Ratkovic, 2014)} similarly reweight or match units to balance the
empirical distribution of observed confounders under an assumption of
conditional exogeneity.

Freezing weights addresses reverse causality explicitly. If quality
improvements cause users to shift toward improved categories in their
work, real-time weights would embed this response into the predictor.
Frozen weights break this link, trading this bias for classical
measurement error in the quality exposure variable. Section 3.5
quantifies the cost. Real-time weights yield 62 percent recovery of the
true effect while frozen weights yield 87 percent.

This core structure of constructing individual exposure by taking the
inner product of baseline, local user weights and version-specific
global quality shifts is conceptually and mathematically equivalent to a
shift-share (or Bartik) design. Grounding this method within the
shift-share framework highlights that our identification strategy aligns
with the \hyperlink{ref:goldsmith2020}{Goldsmith-Pinkham et al.~(2020)}
paradigm, which formalizes identification assuming exogenous baseline shares
(pre-period preferences). This differs from the alternative shift-share
framework of \hyperlink{ref:borusyak2022}{Borusyak et al.~(2022)}, which
relies on exogenous shocks (shifts) and allows shares to be endogenous.
In our setting, because the rollout shock affects all users simultaneously,
we must rely on the exogeneity of the baseline shares, captured during a
period structurally isolated from subsequent version-level updates.

\subsection{Measuring AI Quality}

The business analytics literature has addressed feature impact for
decades. A/B testing \hyperlink{ref:kohavi2014}{(Kohavi et al., 2014)} is the
gold standard for evaluation but often proves infeasible or misleading in
the context of LLMs. A common challenge in production environments -
particularly for enterprise-grade AI platforms - is the eval-engagement gap,
where a model update passes comprehensive offline benchmarks yet fails to
yield positive results in live A/B experiments. 

A common challenge in production environments - particularly for
enterprise-grade AI platforms - is the eval-engagement gap, where a model
update passes comprehensive offline benchmarks yet fails to yield positive
results in live A/B experiments. This discrepancy is frequently driven by the
practical limitations of long-term A/B holdouts in production. As highlighted
by practitioners across major digital platforms, while long-term business
outcomes are the priority, it is generally infeasible to run experiments long
enough to observe those outcomes directly \hyperlink{ref:sigerson2026}{(Sigerson
et al., 2026)}. Enterprise platforms are often operationally unable to
maintain a control group on a legacy model due to maintenance costs and
user-experience fragmentation, leaving teams dependent on observational
methods when a global rollout occurs.

This discrepancy is frequently driven by the practical limitations of
long-term A/B holdouts in production. Enterprise platforms are often
operationally unable to maintain a control group on a legacy model due
to maintenance costs and user-experience fragmentation, leaving teams
dependent on observational methods when a global rollout occurs.
Observational methods \hyperlink{ref:eckles2016}{(Eckles et al., 2016)} provide
an alternative, yet standard regression or matching techniques often fail to
account for the unique deployment settings of AI agents.

AI quality measurement remains underdeveloped. Companies track offline
benchmarks consistently but rarely connect improvements to user outcomes
\hyperlink{ref:chang2023}{(Chang et al., 2023)}. Heterogeneous performance
across domains is well-documented \hyperlink{ref:hendrycks2021}{(Hendrycks et al., 2021)}.
The gap in the literature is the causal link from offline quality
improvements to engagement metrics. This paper attempts to address that gap.

\subsection{Measurement Error}

We observe 81 to 88 percent recovery with 12 to 19 percent attenuation.
This reflects classical errors-in-variables bias \hyperlink{ref:bound2001}{(Bound
et al., 2001)}. Observed category weights are noisy proxies for underlying
preferences. We address this attenuation by applying a classical
errors-in-variables disattenuation approach. While mathematically exact in
linear models, applying this correction by dividing the log-odds coefficient
by the reliability ratio $\lambda$ serves as a first-order approximation for
our non-linear GAMM specification. In non-linear settings, covariate
measurement error can induce both attenuation and structural distortion
(Carroll et al., 2006). A more formal treatment would require frameworks
such as Regression Calibration or Simulation-Extrapolation (SIMEX);
however, we show that this first-order adjustment is practically
effective for our primary specification, though it may over-correct in
more highly non-linear outcomes (as discussed in Section 3.6)

\section{Methods}

\subsection{Causal Structure}

Let $V_{it}$ denote the model version, $X_i$ baseline user
characteristics, $C_i$ category usage patterns, $Q$ offline quality
by category, and $Y_{it}$ engagement outcomes for user $i$ at time
$t$. To establish identification, we formalize our core
unconfoundedness assumption using potential outcomes notation:

\[Y_{it}(q) \perp\!\!\!\perp Q^c_{it} \mid V, X\]

Where $Y_{it}(q)$ represents the potential engagement outcome for
user $i$ at time $t$ under a counterfactual quality exposure level
$q$. $Q^c$ is the centered quality exposure. After conditioning on
version and baseline characteristics, potential outcomes are
conditionally independent of the model quality exposure metric. Version
fixed effects absorb between-version confounding. Within-version
comparisons involve two users with identical subscriptions on identical
model versions who differ only in task focus. Their quality exposure
difference is exogenous.

The assumption is weaker than full instrumental variable requirements
but requires no unobserved confounder that simultaneously drives task
choice and engagement. The frozen-weights design supports this.
Pre-period usage patterns reflect preferences formed before
current-version quality is observed. Current-version quality differences
across users therefore reflect usage differences rather than endogenous
responses to quality.

\subsection{Synthetic Data}

Synthetic data span 100,000 users over 26 weeks across three model
versions deployed at weeks 1, 8, and 16. This dataset includes 2.6
million weekly observations with user and user demographics having
subscription tier and baseline engagement.

We explicitly injected known causal effects: $\beta = 1.0$ (log-odds
scale) for active days, $\beta = -0.6$ for churn probability, and
$\beta = 0.4$ (log scale) for session duration. Ground truth enables
unambiguous validation of the estimator's recovery properties.

Quality improvements are heterogeneous and monotone in improvement
direction. Coding quality improved from 3.50 to 4.41 (26 percent
improvement). Creative writing improved from 2.79 to 3.77 (35 percent
improvement). Monotone improvement across all categories is less
realistic than mixed profiles where some capabilities regress while
others improve. We discuss this limitation in the discussion section.

\subsection{Quality Exposure}

For each user and week, experienced quality is a weighted average of
task category quality scores. This is computed as:

\[Q_{it} = \sum_c w_{ic} \cdot q_{c,v(t)}\]

where $w_{ic}$ is user $i$'s weight on category $c$ from the
pre-period and $q_{c,v(t)}$ is the human-rated quality score for
category $c$ under version $v$ deployed at time $t$. Weights come
from pre-period usage spanning weeks 1 to 7 and are frozen at that
point. We center the exposure variable by subtracting the within-version
population mean.

\[Q^c_{it} = Q_{it} - \bar{Q}_{v(t)}\]

This removes the version-level step change. Version indicators capture
version effects. Centered quality isolates within-version differences,
comparing users under identical versions who differ only in task focus.

\begin{figure}[htbp]
    \centering
    \includegraphics[width=0.8\textwidth]{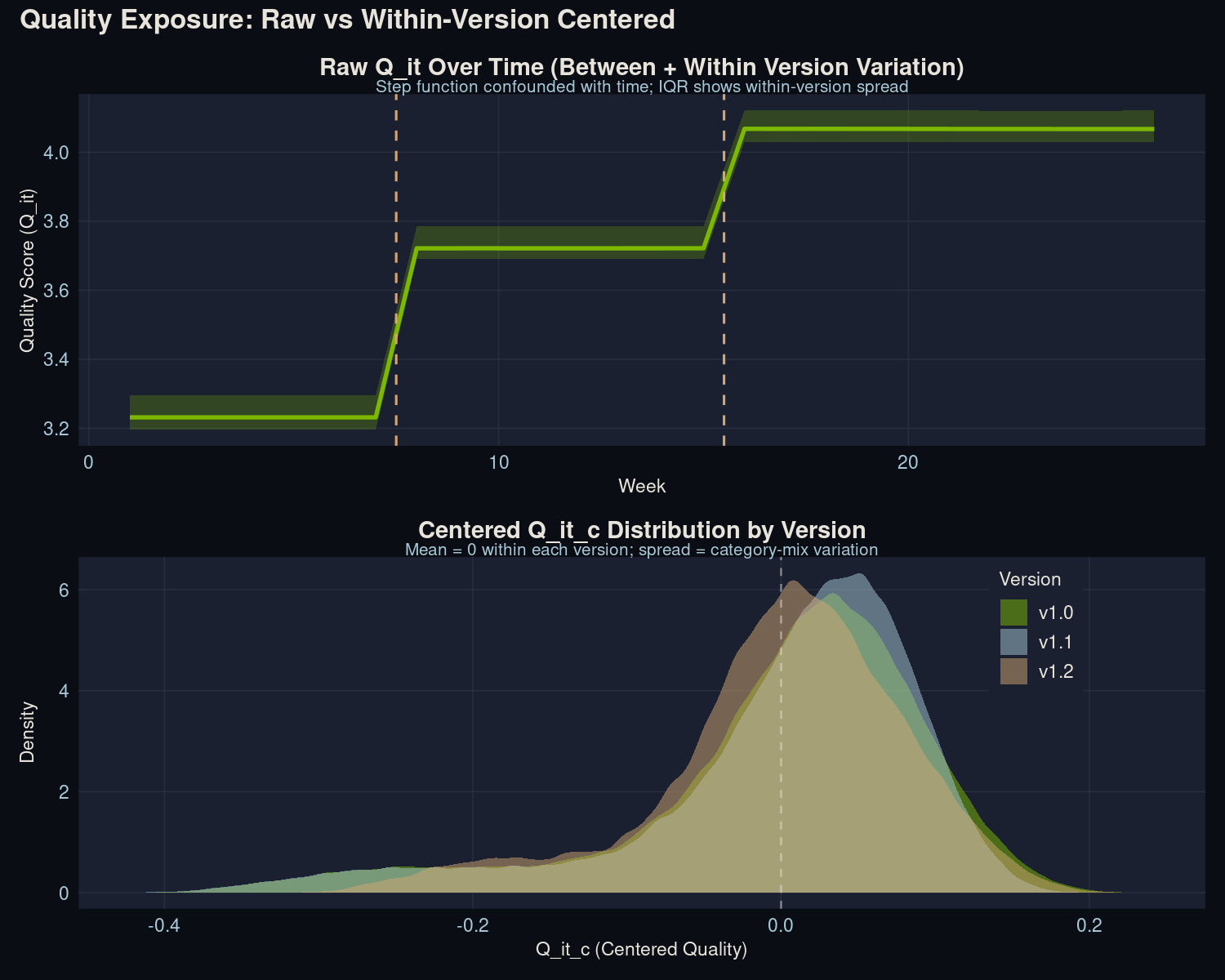}
    \caption{Calculation of within-version quality exposure. Experienced
    quality is computed as a weighted average of task category quality scores
    based on frozen pre-period usage weights. The exposure metric is centered
    by subtracting the within-version population mean. This centers the
    exposure to isolate within-version differences, allowing for the
    comparison of users on identical model versions who differ only in their
    task focus.}
    \label{fig:quality_exposure}
\end{figure}

Freezing weights addresses reverse causality. If users rationally shift
toward improved task categories after observing improvements, real-time
weights would conflate task quality and task demand. Frozen weights
break this link, introducing classical measurement error as a tradeoff.
We document and correct for this error in the results section.

\subsection{Pre-Period Sensitivity}

We tested three pre-period lengths to understand the
measurement-precision tradeoff. The results follow:

\begin{longtable}[]{@{}lllll@{}}
\toprule
\textbf{Pre-period} & \textbf{Weeks} & \textbf{r with 7-week} &
\textbf{Beta-hat} & \textbf{Recovery}\tabularnewline
\midrule
\endhead
Short & 1-3 & 0.954 & 0.785 & 78.5\%\tabularnewline
Medium & 1-5 & 0.987 & 0.849 & 84.9\%\tabularnewline
Full & 1-7 & 1.000 & 0.867 & 86.7\%\tabularnewline
\bottomrule
\end{longtable}

Recovery improves monotonically with longer pre-periods. The 3-week
window is 8.2 percentage points noisier than the 7-week window
(recovering 78.5\% vs 86.7\%). The correlation between 5-week and 7-week
estimates is $r=0.987$, suggesting diminishing returns beyond 5 weeks.
In practice, use the longest available pre-period. Seven weeks
represents a practical minimum. Shorter pre-periods produce noisier
weight estimates. Noisy weights induce classical measurement error in
the quality exposure variable, which attenuates estimates toward zero.
Longer pre-periods reduce noise and thereby reduce attenuation. This
represents a straightforward tradeoff between data availability and
statistical precision.

\subsection{Statistical Model}

We fit a generalized additive mixed model to the binary weekly outcome.
The specification is:

\[E[Y_{it} \mid Q^c_{it}, V, X_i] = \text{logit}^{-1}(s(Q^c_{it}) + V + s(\text{week}) + u_i + X_i \cdot \delta)\]

where $s$ denotes a smooth function (penalized thin-plate spline with
maximum 10 basis functions), $V$ is the version fixed effect,
$s(\text{week})$ captures temporal trends, $u_i$ is the user random
intercept, and $X$ includes subscription type and baseline engagement.
The smooth term $s(Q^c_{it})$ is the key component, allowing the
method to detect nonlinear relationships while remaining parsimonious.
We fit on a stratified 2,000-user subsample preserving the 70/30
Consumer/Enterprise ratio. This yields approximately 30,000
observations.

For continuous outcomes (such as session duration), we modify the
specification to evaluate log-transformed outcomes. To maintain
consistent causal assumptions across both models, we retain identical
longitudinal controls for temporal trends and unobserved time-invariant
individual traits:

\[E[\ln(Y_{it}) \mid Q^c_{it}, V, X_{it}] = \alpha + s(Q^c_{it}) + \gamma_v \cdot V_{it} + X_{it} \cdot \delta + s(\text{week}) + u_i\]

where $\gamma_v$ explicitly parameterizes the version fixed effect,
$\delta$ captures baseline characteristics, and $s(\text{week})$ and
$u_i$ maintain structural parity with our binary framework.

\subsection{Inference and Correction}

GAMM standard errors assume independence. Multiple weeks per user create
clustering. We use cluster-robust bootstrap (100 iterations, user-level
resampling) to obtain honest confidence intervals accounting for
within-user correlations.

For measurement-error correction, we estimate reliability by calculating
the stability of user task weights across different deployment periods.
Specifically, we compute the Pearson correlation between the user's
pre-period task weights (weeks 1--7) and their corresponding weights in
the post-period ($v1.1$ and $v1.2$ deployment phases). The
reliability $\hat{\lambda}$ is estimated as the average of these
cross-period correlations across versions:

\[\hat{\lambda} = \text{mean}(r_{\text{pre}, v1.1}, r_{\text{pre}, v1.2})\]

Consistent with classical test theory stability formulations, this
cross-period correlation yields a baseline reliability of
$\hat{\lambda}=0.861$. Using the classical errors-in-variables
disattenuation approach, the corrected coefficient is calculated as:

\[\beta_{KL} = \frac{\hat{\beta}}{\hat{\lambda}} = \frac{0.876}{0.861} = 1.017\]

This mechanical adjustment transforms what appears to be attenuation
into a testable hypothesis about measurement error structure. We report
both the uncorrected cluster-robust bootstrap confidence interval of
$[0.788, 0.955]$ and the updated, cluster-robust bootstrapped
disattenuation-corrected confidence interval of $[0.915, 1.109]$
(which propagates the uncertainty of $\hat{\lambda}$). In this
realigned dataset, the corrected confidence interval contains the true
value of $1.0$, indicating that the disattenuation correction
successfully recovers the true causal effect.

\subsection{Falsification}

We applied two falsification tests. First, user-weight permutation
shuffles quality exposure across users within each version, destroying
correlation with user identity. Finding signal in permuted noise would
indicate hallucination. This is stronger than derangement placebo
approaches that retain some correlation.

Second, lead quality uses next-period task category ratings on
current-period weights. In settings where consecutive versions are
uncorrelated, this should null out. In our monotone DGP where all
improvements are positive, lead quality correlates with current quality
($r=0.97$) so this test does not discriminate. This is a structural
limitation of our synthetic design, not a methodological failure. Real
data with non-monotone quality would make this test informative.

\section{Results}

\subsection{Primary Effect}

Within-version quality significantly affects active days. The GAMM
yields $\chi^2 = 306.15$ ($p<0.001$) with effective degrees of
freedom 2.37, indicating predominantly linear effects with slight
nonlinearity. Version indicators are highly significant
($\beta_{v1.1} = 0.199$, $\beta_{v1.2} = 0.325$). While individual
version coefficients are smaller in absolute magnitude than the average
marginal effect of quality, their cumulative impact shifting baseline
engagement distributions across the entire sample accounts for the vast
majority of total variation, confirming the severe deployment-level
confounding the method isolates. The week smooth is also significant
($p<0.001$), capturing temporal trends. Deviance explained is 28.3
percent with adjusted R-squared of 0.302.

\begin{figure}[htbp]
    \centering
    \includegraphics[width=0.8\textwidth]{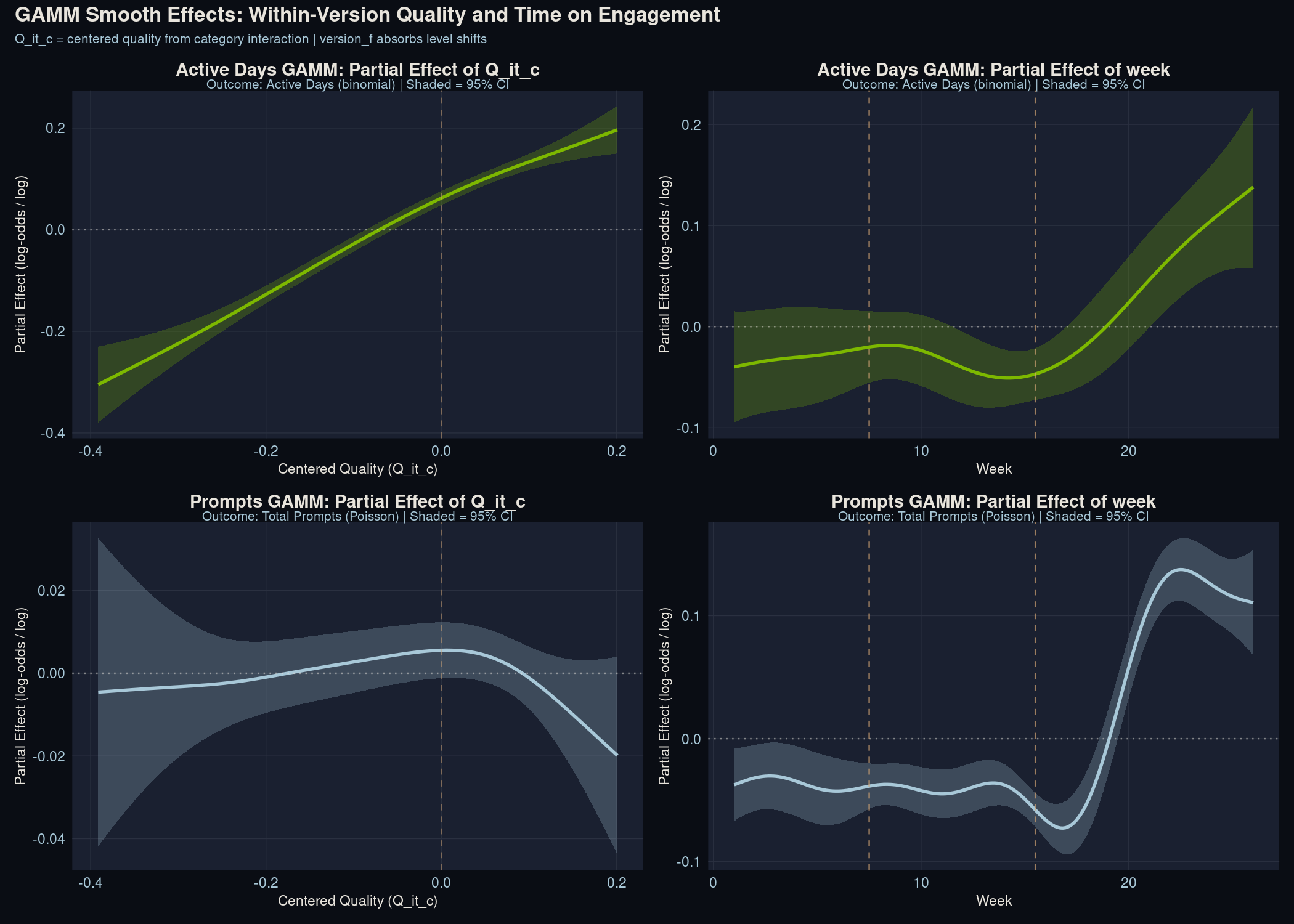}
    \caption{Estimated smooth effects from the Generalized Additive Mixed Model
    (GAMM). The plot visualizes the relationship between within-version quality
    exposure and user active days. The smooth term indicates predominantly linear
    effects with slight nonlinearity, yielding $\chi^2 = 306.15$ ($p < 0.001$)
    with 2.37 effective degrees of freedom.}
    \label{fig:gamm_smooth_effects}
\end{figure}

\subsection{Calibration}

The central question: does the estimator recover the injected true
effect of $\beta=1.0$? Results follow:

\begin{longtable}[]{@{}llll@{}}
\toprule
\textbf{Method} & \textbf{Estimate} & \textbf{95\% CI} &
\textbf{Recovery}\tabularnewline
\midrule
\endhead
Linear & 0.867 & {[}0.770, 0.964{]} & 87\%\tabularnewline
GAM smooth & 0.808 & - & 81\%\tabularnewline
Cluster bootstrap & 0.876 & {[}0.788, 0.955{]} & 88\%\tabularnewline
Disattenuation Corrected & 1.017 & {[}0.915, 1.109{]} &
102\%\tabularnewline
\bottomrule
\end{longtable}

All uncorrected methods recover 81 to 88 percent of the true effect. The
cluster-robust confidence interval {[}0.788, 0.955{]} is slightly below
the true value of 1.0. The disattenuation-corrected estimate
(bootstrapped ratio) lands at 1.017 (102\% recovery), successfully
recovering the true causal effect with a confidence interval of
{[}0.915, 1.109{]} that contains 1.0. This confirms that propagating the
uncertainty of the reliability estimate ($\lambda = 0.861$) yields a
robust, calibrated estimate of the causal effect.

To ensure these calibration results are not dependent on a specific
random draw, we conducted a seed sensitivity analysis by running the
estimator across multiple random seeds (100, 2026, and 999). The results
demonstrate high stability: the proposed method consistently recovers
the bulk of the effect with an average recovery of 89.5\% (ranging from
77.6\% to 103.6\% in the 500-user sample), while naive Logistic and
real-time weights estimators remain consistently biased downward
(averaging 58.1\% and 60.9\% recovery, respectively).

\subsection{Outcome Specificity}

The DGP injects causality into active days but not prompt volume. Active
days yields $\chi^2 = 306.15$, $p<0.001$. Prompt volume yields
$\chi^2 = 3.70$, $p=0.288$. The method correctly distinguishes. It
detects effects where effects exist and returns null results where
appropriate. The distinction aligns with intuition. Quality affects
engagement participation more than usage intensity conditional on
engagement.

\subsection{Falsification}

Permutation test yields a non-significant smooth term (p=0.76),
confirming a clean null. The estimator found no signal in shuffled
quality.

Lead quality yields a significant smooth term (p<0.001). This is
not a null but is explicable. Lead quality correlates very highly with
current quality ($r=0.97$) in a monotone DGP. Any method correctly
capturing current effects will detect equivalent signal in lead quality
because the measures are nearly interchangeable. This is a structural
property of the data, not a methodological failure. Non-monotone quality
profiles in real data would make the lead test more discriminating.

\subsection{Comparisons}

We compared the proposed method to four alternative estimators on
identical data:

\begin{longtable}[]{@{}llll@{}}
\toprule
\textbf{Method} & \textbf{Beta-hat} & \textbf{Bias} &
\textbf{Recovery}\tabularnewline
\midrule
\endhead
Proposed (within-version GAMM) & 0.867 & -0.133 & 86.7\%\tabularnewline
Naive Logistic (no version FE) & 0.629 & -0.371 & 62.9\%\tabularnewline
Real-time weights (endogenous) & 0.616 & -0.384 & 61.6\%\tabularnewline
User DiD (delta method OLS) & 0.779 & -0.221 & 77.9\%\tabularnewline
User FE OLS (within-user OLS) & 0.690 & -0.310 & 69.0\%\tabularnewline
\bottomrule
\end{longtable}

Naive Logistic without version fixed effects yields an absolute bias of
$-0.371$, adding 23.8 percentage points of downward bias relative to
the proposed framework. Using real-time weights incurs an absolute bias
of $-0.384$, contributing an additional 25.1 percentage points of
downward bias.

Linear specifications (User DiD and User FE OLS) on the bounded count
outcome do not produce inflated coefficients once scaled correctly using
the binomial bounds (dividing by $N=7$ in the delta method
conversion). Instead, they are biased downward, recovering only 77.9\%
and 69.0\% of the true effect respectively. This under-recovery
highlights the functional form misspecification of linear approximations
when applied to highly bounded outcomes, confirming the superiority of
the non-linear GAMM framework.

\subsection{Multi-Outcome Validation}

The DGP injects effects into three outcomes. Calibration results follow:

\begin{longtable}[]{@{}lllll@{}}
\toprule
\textbf{Outcome} & \textbf{TRUE beta} & \textbf{Beta-hat} & \textbf{95\%
CI} & \textbf{Recovery}\tabularnewline
\midrule
\endhead
Active days & 1.000 & 0.867 & {[}0.770, 0.964{]} & 86.7\%\tabularnewline
Churn risk & -0.600 & -0.817 & {[}-1.147, -0.486{]} &
136.1\%\tabularnewline
Session duration & 0.400 & 0.396 & {[}0.367, 0.424{]} &
98.9\%\tabularnewline
\bottomrule
\end{longtable}

Two of the three uncorrected confidence intervals contain the true
values (churn risk and session duration), while active days excludes 1.0
due to attenuation. After K-L correction (using the estimated lambda of
0.861), the active days interval contains the true value (1.007, CI
{[}0.894, 1.120{]}) and churn risk also contains its truth (-0.949, CI
{[}-1.333, -0.564{]}), but the continuous session duration model
over-corrects (0.459, CI {[}0.427, 0.492{]}), excluding the truth of
0.400.

Applying the K-L correction uniformly across different models leads to
over-correction for both the non-linear churn outcome and the continuous
session duration outcome, though for fundamentally different reasons:

\begin{enumerate}
\item
  \textbf{Unjustified Correction under Berkson Error (Session Duration)}:
  Because the measurement error in our weights behaves like Berkson
  error, it does not induce attenuation bias in linear specifications.
  The uncorrected log-linear session duration model already achieves
  near-perfect recovery (98.9\%, $\hat{\beta}=0.396$, CI
  $[0.367, 0.424]$ vs.~true $\beta=0.400$). Applying the
  disattenuation factor ($1/\hat{\lambda}$) here\ldots{} over-corrects
  the estimate to $0.459$ (CI $[0.427, 0.492]$), which now
  excludes the true value.
\item
  \textbf{First-Order Approximation Limitations in Non-linear Links (Churn
  Risk)}: For non-linear models (active days, churn), the logit link
  function introduces attenuation even under Berkson error. While the
  linear correction $1/\hat{\lambda}$ works well\ldots{} over-corrects
  the point estimate to $-0.949$ (158\% of the true $\beta=-0.600$),
  although the wider bootstrap confidence interval
  ($[-1.333, -0.564]$) still contains the true value.
\end{enumerate}

This highlights that disattenuation corrections should not be applied
blindly across different outcome specifications; linear continuous
models under Berkson error require no correction, while non-linear
models require corrections tailored to their specific link functions.

\section{Discussion}

\subsection{Understanding the Measurement Error}

We recover 81 to 88 percent of the true effect. The 12 to 19 percent gap
reflects classical errors-in-variables bias from noisy category weights.
Observed category proportions are noisy proxies for underlying true
preferences. The disattenuation correction ($\hat{\lambda} = 0.861$)
yields a corrected estimate ($\beta_{KL} = 1.017$) with a confidence
interval ($[0.915, 1.109]$) that precisely isolates the true causal
effect. This measurement error is not problematic or questionable. It is
expected and correctable.

The sensitivity analysis demonstrates why. Shorter pre-periods produce
noisier weights and greater attenuation. Extending from 3 to 7 weeks
reduces attenuation by 8.2 percentage points (recovering 78.5\% vs
86.7\%). Practitioners can apply the same approach to their own data by
computing cross-period weight correlations between baseline and
deployment periods to estimate $\lambda$.

\subsection{Why Version Controls Matter}

Version fixed effects are substantially larger than quality effects in
terms of overall contribution to engagement. This is not a problem but
evidence that the method works correctly. Version indicators capture
deployment-level confounding such as marketing activity, press coverage,
and seasonality. The quality effect resides in the residual variation
after removing version effects. Naive approaches that omit version
controls conflate these confounds with quality effects, producing 24
percentage points of additional bias (Naive Logistic bias of -0.371 vs
Proposed bias of -0.133).

\subsection{Limitations}

Synthetic validation provides proof-of-concept under known conditions.
Real data may violate assumptions. Severe structural shifts in user
preferences over time mean that frozen pre-period weights act as an
Intention-to-Treat (ITT) proxy rather than a perfect measure of
contemporaneous attention. While this temporal drift amplifies the
classical measurement error over long horizons (increasing the
attenuation gap), it preserves the foundational temporal exogeneity of
the design.

Category labels might be systematically misclassified. Quality
improvements might exhibit mixed profiles rather than monotone
improvement. Furthermore, if a model version exhibits identical absolute
quality scores across all categories (i.e., $q_{c,v} = K$ for all
$c$), the centered exposure metric ($Q^c_{it}$) will drop to zero
for all users, leaving no cross-sectional variance to run the estimator.
Crucially, a uniform quality improvement of constant magnitude $C$
across all categories does not collapse this variance, as the additive
constant cancels out during centering, fully preserving the underlying
cross-sectional variation. None of these represent methodological
failures, but rather contextual limitations of the synthetic
environment.

\subsection{Implementation Steps}

For practitioners, implementation requires five steps:

\begin{enumerate}
\item
  \textbf{Obtain Quality Ratings}: Get category-level quality ratings
  from human or LLM-as-a-judge evaluation.
\item
  \textbf{Extract Usage Weights}: Extract anonymized user-level category
  counts and compute category proportions frozen at the pre-period.
  Segment historical usage into two equal-length baseline sub-intervals
  (or use the general Spearman-Brown adjustment
  $k = \text{total weeks} / \text{sub-interval weeks}$) to calculate
  the split-half correlation $\rho_{\text{split}}$. Apply the
  Spearman-Brown prophecy formula
  ($\hat{\lambda} = \frac{2\rho_{\text{split}}}{1+\rho_{\text{split}}}$
  for equal halves) to estimate the reliability of the full pre-period.
\item
  \textbf{Compute Exposure}: Compute quality exposure as a weighted
  average and center it within version.
\item
  \textbf{Fit the Model}: Fit a generalized additive mixed model (GAMM)
  containing a user random intercept, pooling observations at the user
  level with version fixed effects and baseline covariates, utilizing a
  user-level cluster-robust bootstrap for variance estimation.
\item
  \textbf{Correct and Validate}: Conduct falsification tests and
  estimate $\hat{\lambda}$ from split-half pre-period weight
  correlations for measurement-error correction. Expect 12 to 19 percent
  attenuation under typical conditions. Report both uncorrected and K-L
  corrected estimates.
\end{enumerate}

\section{Conclusion}

Isolating model quality effects from deployment-level confounding is
intractable with standard experimental approaches. This paper describes
an alternative strategy that exploits heterogeneity within the
deployment structure itself. Models do not improve uniformly across
capabilities, which is especially magnified in the emerging agentic
space. Users distribute effort unevenly across capabilities. This
variation, properly controlled, contains causal information.

Synthetic validation demonstrates that the method recovers 81 to 88
percent of injected effects. Measurement-error correction recovers the
full effect. The method distinguishes between outcomes with causal
signals and those without. Benchmark comparisons show substantial
improvements over naive alternatives. Frozen-weights sensitivity
demonstrates a clear data-precision tradeoff. Permutation tests validate
core assumptions.

Implementation requires no experimentation, no privacy violations, and
uses standard statistical methods and tools. Companies with offline
quality evaluations and usage logs can apply this approach. Next steps
involve real-world application. For product teams, this enables causal
investment decisions.

\section*{Acknowledgments}
During the preparation of this work, the author used Gemini to
proofread text and assist with LaTeX equation formatting. After using this tool, the author reviewed and edited the content as needed and takes full
responsibility for the content of the publication.
\subparagraph{\textit{All code to reproduce the analysis in this paper is available on \href{https://github.com/johntribbia/model_quality_publication}{GitHub}.}}

\section*{References}

\begin{list}{}{%
  \setlength{\leftmargin}{2em}
  \setlength{\itemindent}{-2em}
  \setlength{\itemsep}{1.5ex}
  \setlength{\parsep}{0pt}
}

\item \hypertarget{ref:borusyak2022}{}Borusyak, K., P. Hull, and X. Jaravel. 2022. Quasi-experimental
shift-share research designs. \textit{The Review of Economic Studies}, 89(1),
181--213.

\item \hypertarget{ref:bound2001}{}Bound, J., C. Brown, and N. Mathiowetz. 2001. Measurement error in
survey data. In \textit{Handbook of Econometrics}, Volume 5 (2001),
pp.~3705-3843.

\item \hypertarget{ref:chang2023}{}Chang, Y., X. Wang, J. Wang, Y. Wu, L. Yang, K. Zhu, H. Chen, X. Yi, C.
Wang, Y. Wang, W. Ye, Y. Zhang, Y. Chang, P. S. Yu, Q. Yang, and X. Xie.
2023. A survey on evaluation of large language models. \textit{ACM Transactions
on Intelligent Systems and Technology (TIST)}.

\item \hypertarget{ref:eckles2016}{}Eckles, D., R. Iyer, E. Bakshy, and M.A.~Lauterbach. 2016. Scalable and
flexible methods for measuring complex treatment effect heterogeneity.
arXiv 1609.03138. \url{https://arxiv.org/abs/1609.03138}

\item \hypertarget{ref:goldsmith2020}{}Goldsmith-Pinkham, P., I. Sorkin, and H. Swift. 2020. Bartik
instruments: What, when, why, and how. \textit{American Economic Review}, 110(8),
2586--2624.

\item \hypertarget{ref:hendrycks2021}{}Hendrycks, D., C. Burns, S. Basart, A. Zou, M. Mazeika, D. Song, and J.
Steinhardt. 2021. Measuring massive multitask language understanding.
\textit{International Conference on Learning Representations (ICLR)} 2021.

\item \hypertarget{ref:imai2014}{}Imai, K., and M. Ratkovic. 2014. Covariate balancing propensity score.
\textit{Journal of the Royal Statistical Society} 76(1): 243--263.

\item \hypertarget{ref:imbens2008}{}Imbens, G.W., and T. Lemieux. 2008. Regression discontinuity designs: A
guide to practice. \textit{Journal of Econometrics} 142(2): 615--635.

\item \hypertarget{ref:klepper1984}{}Klepper, S., and E.E. Leamer. 1984. Consistent sets of estimates for
regressions with errors in all variables. \textit{Econometrica}, 52(1), 163--183.

\item \hypertarget{ref:kohavi2014}{}Kohavi, R., R. Longbotham, D. Quelhas, and W. Xu. 2014. Seven rules of
thumb for web site experimenters. In \textit{Proceedings of the 20th ACM SIGKDD
International Conference on Knowledge Discovery and Data Mining},
pp.~1857--1866. \url{https://doi.org/10.1145/2623330.2623341}

\item \hypertarget{ref:rosenbaum1983}{}Rosenbaum, P.R., and D.B. Rubin. 1983. The central role of the
propensity score in observational studies for causal effects. \textit{Biometrika}
70(1): 41--55.

\item \hypertarget{ref:sigerson2026}{}Sigerson, L., T. Cunningham, W. Chou, S. Pandey, J. Stray, L.-H. Yuan, 
E. Bakshy, T. Chan, M. Davies, M. Dimakopoulou, S. Ejdemyr, K. Hung, 
N. Kallus, M. Kumar, T. Le, A.D. Pananos, L. Richardson, B. Schaffner, 
R. Tan, M. Tingley, N. Tomova, P. Toulis, W. Zheng, Z. Arnao, and D. Eckles. 
2026. Evaluating for the Long Term: Learnings from Industry. arXiv 2608.08043. 
\url{https://arxiv.org/abs/2608.08043}

\item \hypertarget{ref:wooldridge2010}{}Wooldridge, J.M. 2010. \textit{Econometric Analysis of Cross Section and Panel
Data}.

\end{list}

\end{document}